\documentclass[prc,reprint,longbibliography,superscriptaddress,amsmath]%
{revtex4-1}

\usepackage{color}
\definecolor{goodgreen}{rgb}{0.1,0.5,0}
\definecolor{goodred}{rgb}{0.7,0,0}
\definecolor{goodblue}{rgb}{0,0,0.8}
\usepackage[colorlinks,urlcolor=blue,citecolor=goodgreen,linkcolor=goodred]%
{hyperref}
\usepackage{graphicx}
\usepackage[utf8]{inputenc}
\usepackage{mathptmx}
\usepackage[scaled=.90]{helvet}
\usepackage{courier}

\newcommand{\un}[1]{\ensuremath{\,\text{#1}}}
\newcommand{\bpar}{\ensuremath{B_{||}}}
\newcommand{\vg}{\ensuremath{V_\text{g}}}
\newcommand{\vsd}{\ensuremath{V_\text{sd}}}

\begin{document}

\title{From Transparent Conduction to Coulomb Blockade at Fixed Hole Number}

\author{D. R. Schmid}
\affiliation{Institute for Experimental and Applied Physics, 
University of Regensburg, 93040 Regensburg, Germany}
\author{P. L. Stiller}
\affiliation{Institute for Experimental and Applied Physics, 
University of Regensburg, 93040 Regensburg, Germany}
\author{A. Dirnaichner}
\affiliation{Institute for Experimental and Applied Physics, 
University of Regensburg, 93040 Regensburg, Germany}
\author{A. K. Hüttel}
\email{andreas.huettel@ur.de}
\affiliation{Institute for Experimental and Applied Physics, 
University of Regensburg, 93040 Regensburg, Germany}
\affiliation{Low Temperature Laboratory, Department of Applied Physics, 
Aalto University, P.O. Box 15100, 00076 Aalto, Finland}

\date{September 20, 2020}

\begin{abstract}
We present a complex set of transport spectroscopy data on a clean single-wall 
carbon nanotube device in high magnetic fields. At zero axial field, the device 
displays in hole conduction with increasingly negative gate voltage a fast 
transition towards high contact transparency and eventually Fabry-P\'erot 
interference of conductance. When increasing the axial field component up to 
$\bpar=17\un{T}$, the contact transparency and the overall conductance are 
reduced all the way to Coulomb blockade, clearly displaying the subsequent 
charging with the first 10 holes. The continuous transition between the 
transport regimes is dominated by a rich spectrum of Kondo-like resonances, with 
distinct features in the stability diagrams.
\end{abstract}

\maketitle

\section{Introduction}

Carbon nanotubes provide a prototypical, highly versatile system for quantum 
transport, which has been the topic of extensive research over the last decades 
\cite{rmp-laird-2015}. They have allowed the observation of electronic 
transport regimes as different as Coulomb blockade 
\cite{nature-tans-1997,highfield}, Kondo effect-dominated tunneling 
\cite{nature-nygard-2000,kondocharge}, and electronic Fabry-P\'erot 
interference \cite{nature-liang-2001,fabryperot}. A striking property of 
suspended, so-called ``ultraclean'' nanotube devices 
\cite{highfield,nmat-cao-2005} is that the effective transparency of contacts 
can be tuned over a large range by application of a gate voltage alone, while 
maintaining the regularity of the confinement potential. In combination with 
analysis of the repetitive shell filling, this has led to studies on the 
evolution of transport regimes with 
tunnel coupling \cite{prl-makarovski-2007,prl-anders-2008,arxiv-yang-2020}.

When the absolute number of electrons or holes, as opposed to the shell 
filling, is relevant, investigating the dependence of the spectrum on barrier 
transparencies becomes more challenging. Approaches that have been pursued here 
include comparing hole and electron spectrum \cite{natcomm-niklas-2016} (which 
utilizes electron-hole symmetry) or introducing additional barrier gates 
\cite{nphys-benyamini-2014} (which requires more complex fabrication).

Here, we present data on the transport spectrum in the few-hole regime, where 
the contacts are transparent and typically a transition to Fabry-P\'erot 
interference of conductance is observed \cite{nature-liang-2001,fabryperot}. 
Application of an axial magnetic field of up to $\bpar=17\un{T}$ reduces the 
conductance in our device, leading via multiple Kondo-like transport resonances
\cite{nature-nygard-2000,kondocharge,nature-goldhabergordon-1998,%
nature-jarillo-2005,brokensu4} all the way to 
strong Coulomb blockade, at a then known number of holes in the valence band. A 
comprehensive theoretical model for the data is so far still missing, however, 
we hope to inspire corresponding work. In order to facilitate this, the raw 
data of the measurements presented here is deposited online under an open 
access license \cite{zenodo}.

\section{Device and measurement}

\begin{figure*}[thb]
\includegraphics[width=\textwidth]{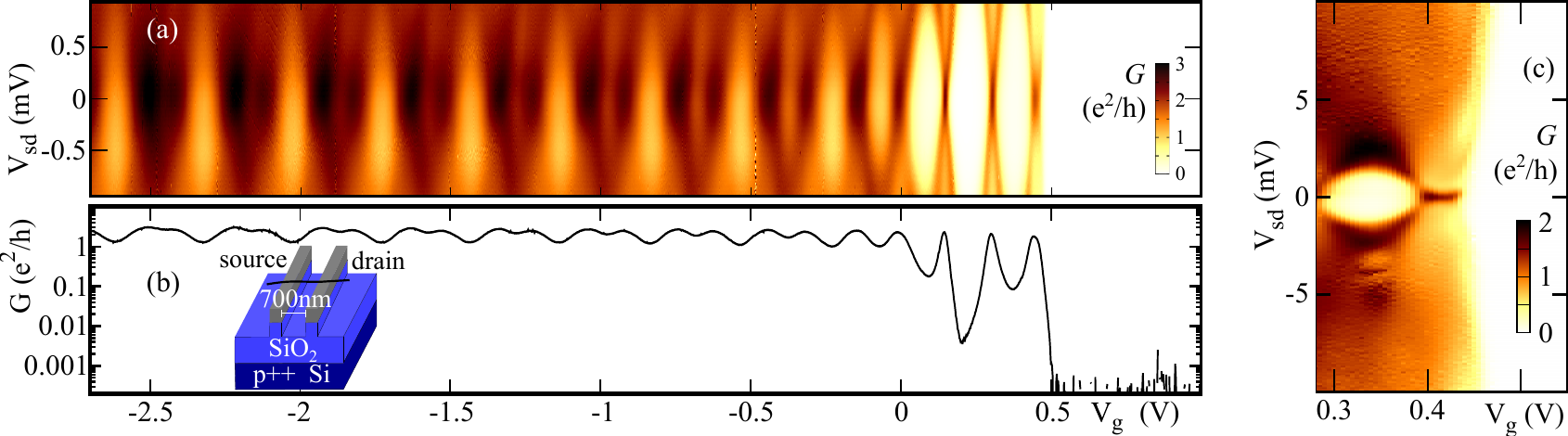}
\caption{%
Inset: Sketch (not to scale) of the central part of the nanotube device, with a 
suspended carbon nanotube grown in situ across rhenium contacts. 
(a) Differential conductance $G(\vsd, \vg)$ in the few-hole region, 
numerically differentiated from a dc current measurement; pre-characterization 
measurement at $T\simeq 300\un{mK}$.
(b) Zero dc bias conductance $G(\vg)$ as function of gate voltage \vg;
trace cut from the data of (a).
(c) Differential conductance $G(\vsd, \vg)$ near the band gap edge, from a 
detail dc current measurement for increased bias range and gate resolution, 
$T\simeq 300\un{mK}$.
}
\label{fig1}
\end{figure*}
The inset of Figure~\ref{fig1} displays a schematic of the measured device. 
Using chemical vapor deposition (CVD) \cite{nature-kong-1998}, a carbon 
nanotube has been grown {\it in situ} across predefined, 40\,nm thick rhenium 
contacts and a trench between them; the highly doped silicon chip substrate is 
used as a global back gate. The contact distance, as approximation for the 
suspended length of the nanotube segment, is $L=700\un{nm}$. During the CVD 
process, the contact metal surface is saturated with carbon \cite{remo}; barely 
any traces of superconductivity in the contacts can be found in the transport 
measurements.

The device was very stable, surviving multiple cool-downs in different 
cryostats with only minor changes, and very clean in the sense of highly 
regular transport spectra. Its characteristics in parameter ranges other than 
discussed here have already been presented in several publications 
\cite{highfield,kondocharge,magdamping,heliumdamping,franckcondon}. The device 
shows the typical behaviour of a small-bandgap carbon nanotube. For positive 
gate voltages $\vg > 0.6\un{V}$, i.e., in electron conductance, strong Coulomb 
blockade with a repetitive shell filling pattern can be observed (not shown); 
the tunnel barriers here are formed by a wide p-n junction within the carbon 
nanotube \cite{highfield,apl-park-2001,nature-kuemmeth-2008}. The electronic 
band gap is observed at a low positive gate voltage.

Figures~\ref{fig1}(a--c) show pre-characterization measurements of hole 
transport, for $\vg < 0.5\un{V}$, performed in the vacuum of a helium-3 
cryostat at $T\simeq 300\un{mK}$. The differential conductance $G(\vg, 
\vsd)$ as function of gate voltage \vg\ and bias voltage \vsd\ is plotted in 
Fig.~\ref{fig1}(a). While at this temperature close to the band gap edge 
Coulomb blockade related effects are still visible, the transition into 
electronic Fabry-P\'erot interference with a strongly broadened, oscillatory 
pattern and a conductance always exceeding $G_0=e^2/h$ \cite{arxiv-yang-2020} 
is clear for $\vg \ll 0\un{V}$. This is also demonstrated by the zero bias 
conductance trace plotted in Fig.~\ref{fig1}(b), which shows a behaviour 
strongly resembling, e.g., data published in \cite{fabryperot}, including the 
onset of a Sagnac interference-induced modulation.

On closer observation, the precise nature of the conductance oscillations near 
the electronic band gap already turns out to be more complex than expected. 
Figure~\ref{fig1}(c) shows a detail measurement of the rapid conductance onset 
at the gap edge. The measurement exhibits no indications of further structure 
within the gap region even at high bias, strong filtering and wide-range 
logarithmic color scale plotting, indicating that the nanotube is here fully 
depleted of free carriers. The first electron enters the conductance band at 
$\vg\simeq 0.68\un{V}$ with a sharp Coulomb oscillation (not shown). Conversely, 
the low-bias conductance maximum near $\vg = 0.4\un{V}$ resolves into a 
structure broadly extended in \vg, allowing for speculation that it represents 
a Kondo ridge with already the addition of two holes 
\cite{nature-nygard-2000,kondocharge}.

\section{Large axial magnetic field}

In the following, all presented data has been recorded at base temperature 
$T\simeq 30\un{mK}$ of a top-loading dilution refrigerator, with the device 
immersed into the diluted phase of the liquid $^3$He/$^4$He mixture. The 
dilution refrigerator was equipped with a $17\un{T}$ superconducting magnet and 
a rotateable sample holder, such that the relative orientation of the magnetic 
field and the carbon nanotube could be adjusted within the chip surface 
plane.

\begin{figure}[t]
\includegraphics[width=\columnwidth]{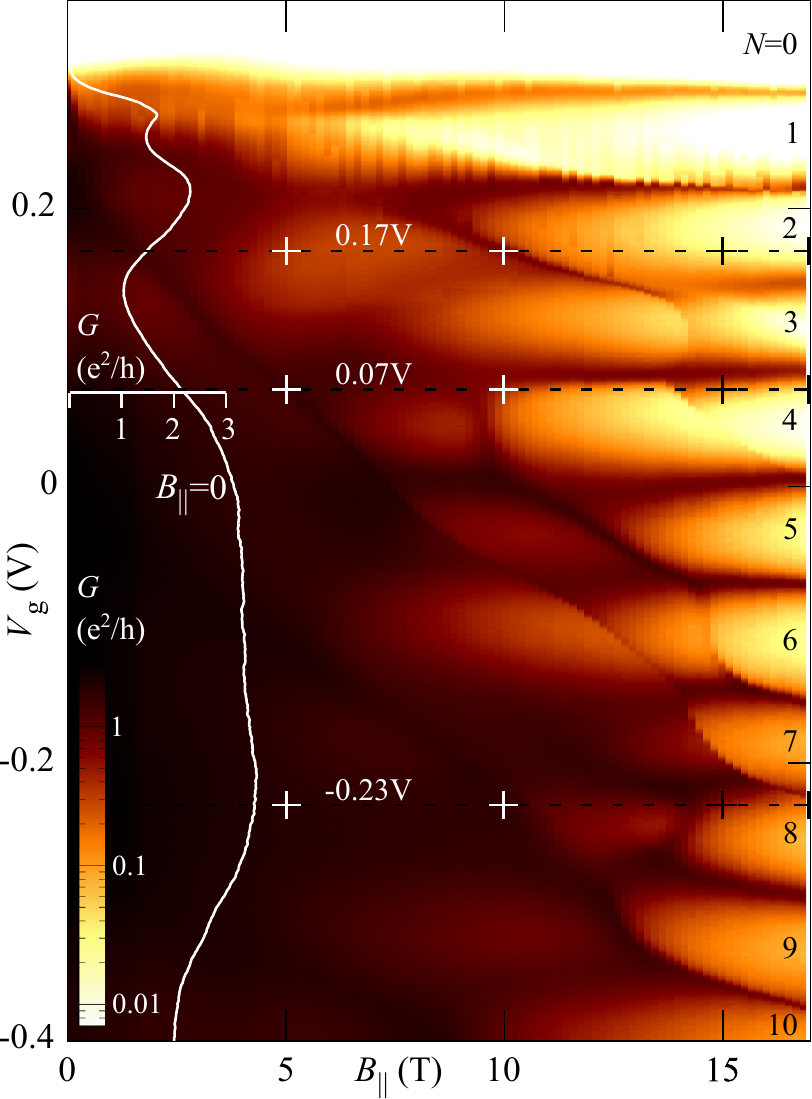}
\caption{
Zero bias conductance $G(\bpar, \vg)$ in the few-hole regime at dilution 
refrigerator base temperature $T \simeq 30\un{mK}$, as function of gate voltage 
\vg\ and magnetic field \bpar\ parallel to the carbon nanotube axis. Absolute 
hole numbers at large field are indicated by numbers at the right graph edge. 
The overlaid line trace plots the conductance at $\bpar=0$ (same \vg\ axis). 
Dashed lines and crosses mark the parameters of the trace cuts in 
Fig.~\ref{fig4}.
}
\label{fig2}
\end{figure}
Our central observation is shown in Fig.~\ref{fig2}. It plots the zero-bias 
conductance $G(\bpar,\vg)$ in the few-hole regime, as function of both gate 
voltage \vg\ and magnetic field in the direction of the carbon nanotube axis 
\bpar, over a wide field range $0\un{T} \le \bpar \le 17\un{T}$.

At low magnetic field, left edge of the plot, the transition towards
Fabry-Pérot interference \cite{nature-liang-2001,fabryperot} with an overall 
increasing conductance is visible. For further clarity, the conductance trace 
$G(0\un{T}, \vg)$ of the data set has been overlaid as a white line plot; $G$ 
exceeds $3 e^2/h$ around $\vg=-0.2\un{V}$. It shows several oscillations; 
comparing the gate voltage scale of the broad ridge in the region $-0.3\un{V} 
\le \vg \le 0\un{V}$ with the high field behaviour discussed below and with 
Fig.~\ref{fig1} indicates that we observe here already a $4e$-periodic 
phenomenon, at the transition between SU(4) Kondo effect and Fabry-Pérot 
interference \cite{arxiv-yang-2020}.

Conversely, at high axial magnetic field, up to $\bpar = 17\un{T}$ at the 
right edge of the plot, the overall conductance is significantly lower, and the 
nanotube device displays well-separated Coulomb oscillations of conductance. A 
careful search at the edge of the band gap region has not shown any indications 
of further features in hole conductance at more positive \vg. This allows to
identify the absolute number of holes $N$ in the system in Coulomb blockade, 
see the numbers at the right edge of the plot.

The transition from transparent conductance to Coulomb blockade behaviour, 
filling the central area of the plot, displays an extraordinary degree of 
complexity. Kondo-like resonances cross Coulomb blockade areas, with different 
levels of background (cotunneling) conductance on either side of them. They 
merge into diagonal features in the plot, for lower field passing to more 
positive gate voltages. From a theory perspective, the transition has to 
involve a reduction of the charging energy per hole from being the dominant 
energy scale to nearly zero, giving an indication of the expected multitude of 
phenomena.

\begin{figure*}[thb]
\includegraphics{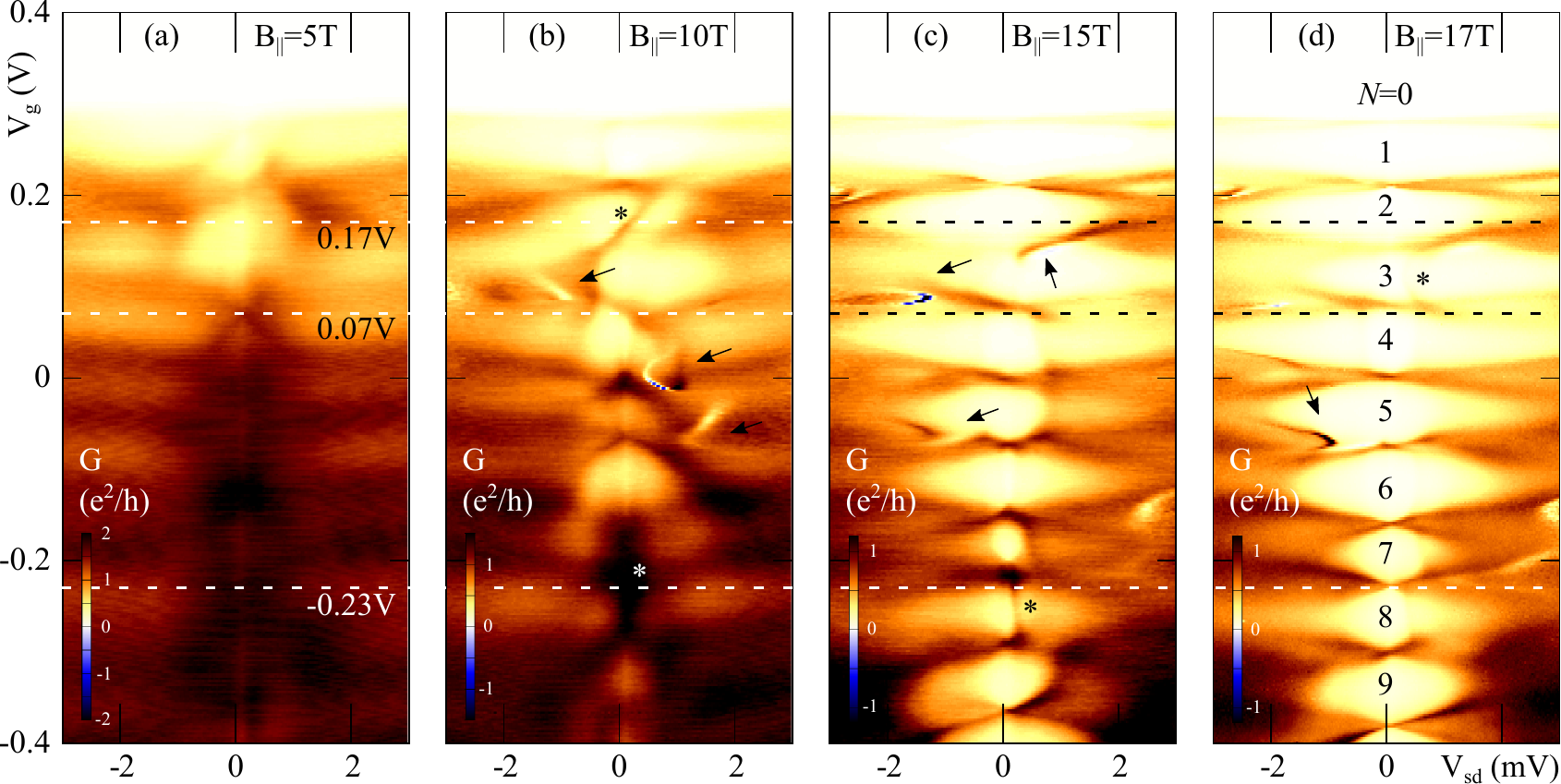}
\caption{%
Stability diagrams, plotting the differential conductance $G(\vsd, \vg)$ as 
function of bias voltage \vsd\ and gate voltage \vg, for fixed magnetic field 
parallel to the cabon nanotube axis: (a) $\bpar=5\un{T}$, (b) $\bpar=10\un{T}$, 
(c) $\bpar=15\un{T}$, (d) $\bpar=17\un{T}$. Negative differential conductance 
is plotted in blue. The dashed lines mark the gate voltage values of the trace 
cuts in Fig.~\ref{fig4}. Asterisks indicate Kondo-like low-bias anomalies of 
conductance, arrows sickle-shaped features, see the text.
}
\label{fig3}
\end{figure*}
Extending the data of Fig.~\ref{fig2}, Fig.~\ref{fig3} plots stability 
diagrams, i.e., the differential conductance $G(\vsd, \vg)$ as function of bias 
voltage \vsd\ and gate voltage \vg, for fixed values of the magnetic field 
\bpar\ in each panel. At comparatively low magnetic field $\bpar=5\un{T}$, only 
strongly broadened patterns can be observed, see Fig.~\ref{fig3}(a). At 
$\bpar=10\un{T}$, Fig.~\ref{fig3}(b), first features resembling diamond-shaped 
Coulomb blockade regions emerge. They are strongly distorted and broadened, 
overlaid by a wide zero-bias ridge of enhanced conductance at $\vg \simeq 
-0.2\un{V}$, and by strongly bias-dependent resonances as, e.g., at $\vg \simeq 
0.17\un{V}$. The trend towards lower conductance and more regular Columb 
blockade regions continues with $\bpar=15\un{T}$, Fig.~\ref{fig3}(c), where 
except for the region near $\vg\simeq -0.25\un{V}$ the zero-bias conductance 
anomalies are nearly gone, and $\bpar=17\un{T}$, Fig.~\ref{fig3}(d).

\section{Detail observations}
\begin{figure}[t]
\includegraphics{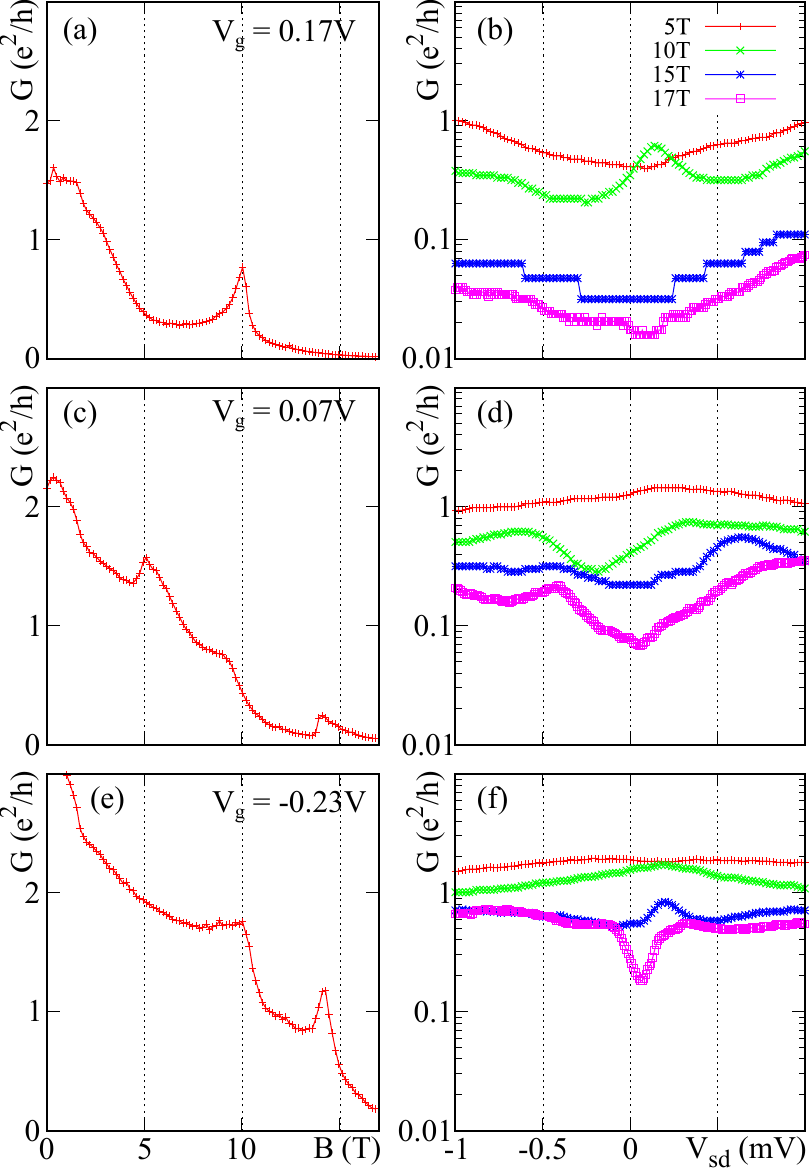}
\caption{%
Conductance traces for fixed gate voltage (a,b) $\vg=0.17\un{V}$, (c,d) 
$\vg=0.07\un{V}$, and (e,f) $\vg=-0.23\un{V}$. (a,c,e) are trace cuts from 
Fig.~\ref{fig2}, plotting the differential conductance $G(\bpar)$ as function 
of parallel magnetic field, at constant $\vsd\simeq 0$. (b,d,f) are the
corresponding sections of the panels of Fig.~\ref{fig3}, plotting the 
conductance $G(\vsd)$ as function of bias, for constant $\bpar=5, 10, 15, 
17\un{T}$. In (b), the lock-in amplifier reached its resolution limit in the 
$15\un{T}$ trace, leading to step-like measurement artifacts.
}
\label{fig4}
\end{figure}
To gain insight into the nature of the ``diagonally running resonances'' in the 
magnetoconductance spectrum of Fig.~\ref{fig2}, in Fig.~\ref{fig4} we plot 
selected line traces both from Fig.~\ref{fig2} and Fig.~\ref{fig3}. We chose 
gate voltages such that at the magnetic fields $\bpar$ of the stability 
diagrams of Fig.~\ref{fig3} such a resonance is observed. The corresponding 
gate voltages are marked in Fig.~\ref{fig2} and Fig.~\ref{fig3} with dashed 
lines.

For the first two panels, Fig.~\ref{fig4}(a,b), the gate voltage is 
$\vg=0.17\un{V}$; at this voltage, a resonance crosses $\bpar=10\un{T}$, see 
the white cross in Fig.~\ref{fig2}. This resonance becomes clearly visible 
again in the line cut of Fig.~\ref{fig4}(a) as a distinct local maximum of 
conductance. Fig.~\ref{fig4}(b) shows the corresponding bias traces of the 
panels of Fig.~\ref{fig3}, at $\bpar=5, 10, 15, 17\un{T}$. The trace at 
$\bpar=10\un{T}$ immediately stands out with a (near) zero bias anomaly of 
conductance, suggesting a Kondo-like behaviour of the observed phenomenon.
Surprisingly, the conductance maximum displays a strong gate voltage 
dependence in Fig.~\ref{fig3}(b).

Similar behaviour of the line traces is observed in Fig.~\ref{fig4}(c,d) at 
$\bpar=5\un{T}$ and in Fig.~\ref{fig4}(e,f) at $\bpar=10\un{T}$. For 
Fig.~\ref{fig4}(c,d), $\bpar=5\un{T}$ and $\vg\simeq 0.07\un{V}$, the 
corresponding region in the stability diagram of Fig.~\ref{fig3}(a) resembles a 
Coulomb blockade degeneracy point; it also evolves at higher magnetic field 
into the $3\le N \le 4$ transition. In the case of Fig.~\ref{fig4}(e,f) at 
$\bpar=10\un{T}$ and $\vg\simeq -0.23\un{V}$, the zero-bias conductance maximum 
corresponds to a wide region in Fig.~\ref{fig3}(b) similar to merged Kondo 
ridges. In addition, in the latter panels, also the conductance feature near 
but not exactly at $\bpar=15\un{T}$ translates into a conductance maximum in 
bias dependence. Note that the data indicates a global shift of all features in 
bias on the order of $\Delta\vsd\sim +0.1\un{meV}$, most likely due to an input 
offset of the current amplifier. On close observation the same offset is also 
visible in Fig.~\ref{fig3}.

\begin{figure}[t]
\includegraphics{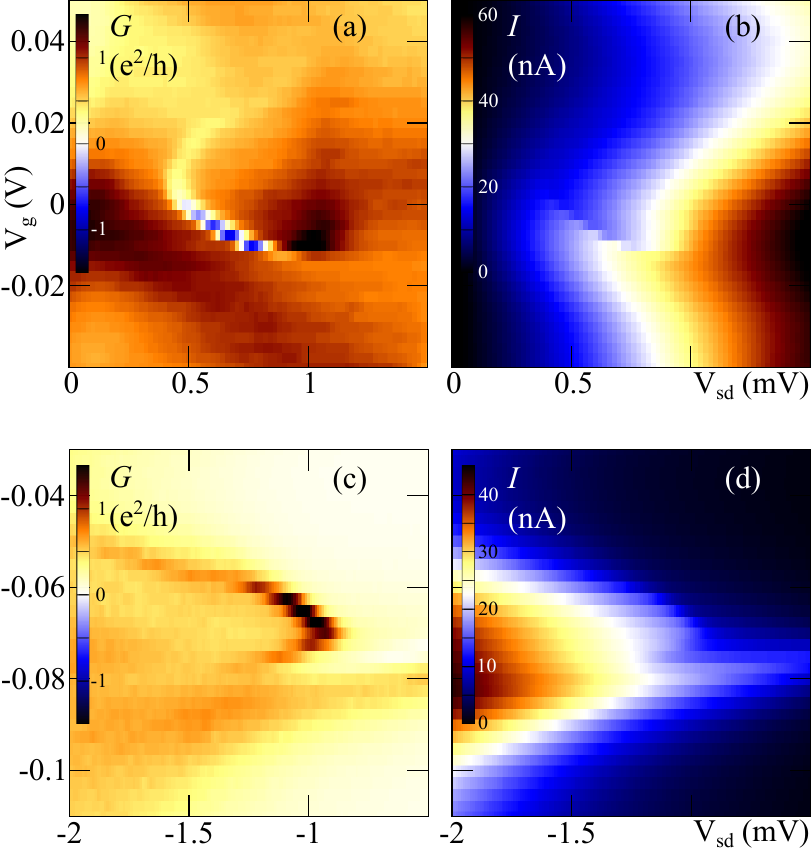}
\caption{%
(a) Detail enlargement of the differential conductance measurement 
$G(\vsd, \vg)$ of Fig.~\ref{fig3}(b); $\bpar=10\un{T}$. (b) Simultaneously 
measured dc current $I(\vsd, \vg)$. (c) Detail enlargement of the differential 
conductance measurement $G(\vsd, \vg)$ of Fig.~\ref{fig3}(d); 
$\bpar=17\un{T}$. (d) Simultaneously measured dc current $I(\vsd, \vg)$.
}
\label{fig5}
\end{figure}
A surprising detail of the stability diagrams of Fig.~\ref{fig3} is the 
presence of several sickle-shaped features of positive, low, or even strong 
negative differential conductance, see the arrows in the figure. Two of the 
corresponding regions are enlarged in Fig.~\ref{fig5}, each accompanied by the 
simultaneously measured dc current. 

The precise origin of these features is unknown; their shape does not 
correspond to typical single electron tunneling or cotunnelling phenomena.
Given that the measurement took place with the device immersed in liquid 
helium, and that a high magnetic field was present, mechanical self-activation 
is unlikely to be the cause
\cite{magdamping,heliumdamping,strongcoupling,nphys-wen-2020,nphys-urgell-2020}%
. A self-driving mechanism would need to overcome both viscous damping and 
dissipation due to induction. An unambiguous decision whether the sharpest such 
features here in our measurement are sudden switching events, as expected for 
the onset of mechanical instability, is not possible from the data set due to 
the averaging times of lock-in amplifier and multimeter. Some of the 
sickle-shaped features in Fig.~\ref{fig3} are, however, clearly broadened, see, 
e.g., the region enlarged in Fig.~\ref{fig5}(c,d), also speaking against a 
vibrational instability phenomenon.

\section{Further data}

\begin{figure}[t]
\includegraphics{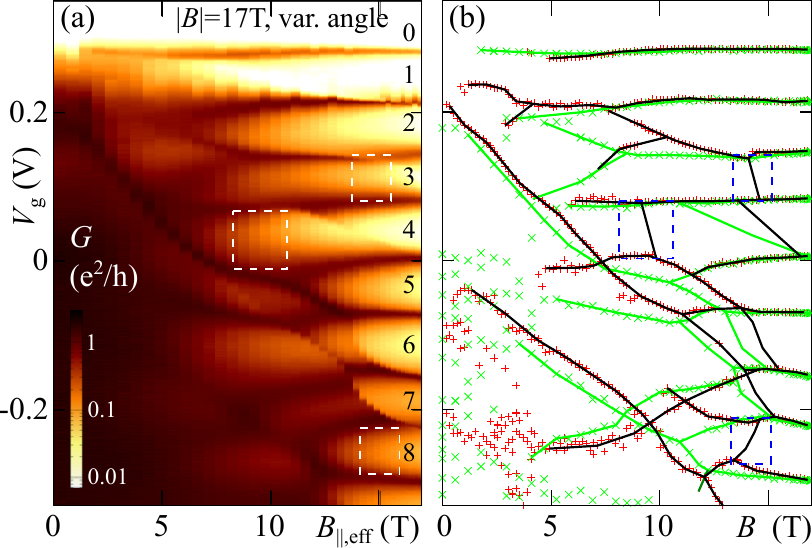}
\caption{%
(a) Zero bias conductance $G(B_{||,\text{eff}}, \vg)$ as function of gate 
voltage \vg\ and effective axial magnetic field $B_{||,\text{eff}} = B \cos 
(\phi)$, for constant $B=17\un{T}$  and varied angle $\phi$ between nanotube 
axis and field direction. The dashed white rectangles mark regions where the 
pattern clearly deviates from Fig.~\ref{fig2}.
(b) Points: automatically extracted local maxima of the $G(\vg)$ traces of the 
datasets of Fig.~\ref{fig2} and Fig.~\ref{fig6}(a); solid lines: manually 
inserted guides to the eye, also taking into account step-like features. Red 
points / black lines: varied parallel field, data of Fig.~\ref{fig2}; green 
points / green lines: rotating device in constant field, data of 
Fig.~\ref{fig6}(a).
}
\label{fig6}
\end{figure}
In Fig.~\ref{fig6}(a), the magnetic field is kept constant at its maximum 
value, $B=17\un{T}$, but the sample holder is rotated stepwise; the magnetic 
field remains in the chip surface plane, but the relative orientation of 
nanotube axis and field changes. The plot shows the differential conductance as 
function of gate voltage and axial magnetic field component $B_{||,\text{eff}} 
= B \cos (\phi)$, with $\phi$ as the angle between field direction and nanotube 
axis. The overall similarity between Fig.~\ref{fig6}(a) and Fig.~\ref{fig2} 
immediately confirms that the axial field component is crucial for the 
suppression of overall conductance and for most of the observed spectrum 
features. On closer observation, however, there are subtle deviations in the 
conductance resonance pattern between the figures, indicating that the large 
angle-independent Zeeman energy here modifies the transport spectrum. Three 
particularly clear cases where resonance lines visible in Fig.~\ref{fig2} are 
absent in Fig.~\ref{fig6}(a) are marked with dashed rectangles in the figure.

Figure~\ref{fig6}(b) explicitly compares the two measurements. Local maxima 
have been extracted from $G(\vg)$ traces and plotted as points; based on these 
points, lines have been drawn manually to connect them, also taking account 
step-like features in the data that did not trigger the maximum search. As last 
step, the two sets of points and lines have been superimposed. Red points and 
black lines stem from the data of Fig.~\ref{fig2}, where an axial magnetic 
field of varied strength is applied; green points and green lines stem from 
Fig.~\ref{fig6}(a), where at fixed magnetic field value $B=17\un{T}$ the angle 
$\phi$ between nanotube and field and with that the axial component is varied. 
As expected, at the right edge of the plot, with $\bpar=17\un{T}$ (or $\phi=0$) 
the patterns coincide. For smaller field or larger angle, the deviations 
gradually increase; the three resonances missing in Fig.~\ref{fig6}(a) are 
marked again with dashed lines.

\begin{figure}[t]
\includegraphics{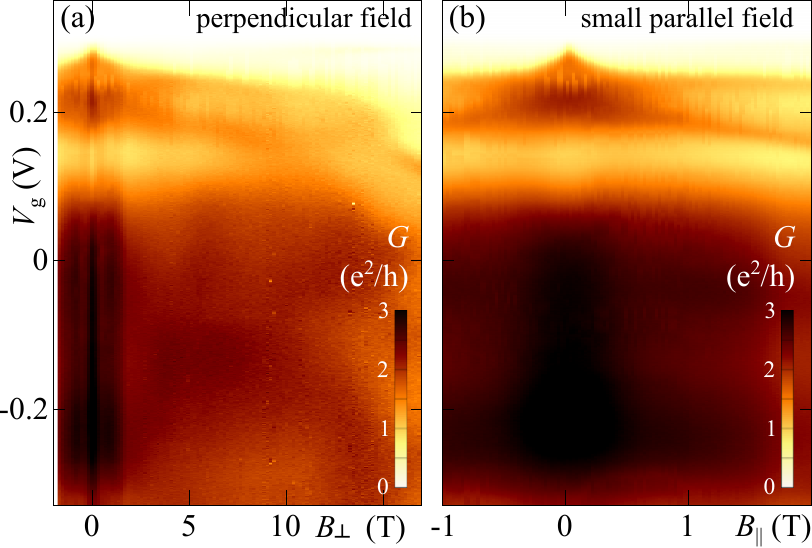}
\caption{%
(a) Conductance $G(B_\perp, \vg)$ as function of gate voltage \vg\ and magnetic 
field $B_\perp$ perpendicular to the nanotube axis.
(b) Conductance $G(\bpar, \vg)$ as function of gate voltage \vg\ and magnetic 
field \bpar\ parallel to the nanotube axis, in higher resolution for the 
low-field range $-1\un{T}\le \bpar \le 2\un{T}$.
}
\label{fig7}
\end{figure}

Figure~\ref{fig7}(a) plots the conductance in the case of a magnetic field 
perpendicular to the nanotube axis, $G(B_\perp, \vg)$. The overall signal 
remains large, and only broad features occur. Around zero magnetic field, a 
gate voltage independent pattern of larger conductance is visible, which may be 
related to superconductivity in the rhenium contacts.

Finally, Fig.~\ref{fig7}(b) shows a detail measurement at low parallel magnetic 
field $-1\un{T} \le \bpar \le 2\un{T}$. Again, around zero field, larger 
conductance is observed, consistent with an impact of the contact metal 
independent of the precise in-plane orientation of the magnetic field.

\section{Discussion}

While significant research has been published regarding carbon nanotubes grown
in-situ over metal contacts, the precise interface properties and contact 
geometries in low-temperature experiments still pose open questions. In the 
case of electron transport, quantum dot confinement is typically given by 
an electrostatically induced p-n junction within the nanotube, and the contacts 
to the quantum dot are formed by segments of the nanotube itself beyond these 
junctions. The nanotube-metal interface can be treated as small perturbation or 
even neglected \cite{highfield,apl-park-2001}. This is, however, not the case 
for hole transport (i.e., negative applied gate voltages), where the p-n 
junction is absent. 

Regarding the overall suppression of conductance with increasing axial magnetic 
field, it has been demonstrated recently that cross-quantization affects in 
such a field the shape of electronic wave functions along the carbon nanotube 
axis \cite{highfield,franckcondon}. This leads to smaller tunnel rates to the 
contacts in the high field limit. The observation in \cite{highfield} and the 
corresponding theoretical analysis was, however, targeted at the case of single 
electron states in a well-closed off system. It may require extension or 
modification for transparent contacts and interacting charges. 

In \cite{prl-groverasmussen-2012}, also an impact of a magnetic field on the 
tunnel coupling is discussed, though state-specific for KK'-mixed doublet states 
and on a smaller magnetic field scale. The mechanism proposed in 
\cite{prl-groverasmussen-2012} is that the axial field modifies the 
circumferential wave function and with that the wavefunction overlap at a side 
contact. Given that in our device the nanotube lies on top of the contacts, 
this may well be relevant specifically in the hole regime.

Regarding carbon nanotube transport spectra in magnetic fields, many works have 
demonstrated complex results including multiple ground state transitions as 
well as, e.g., Kondo phenomena caused by nontrivial degeneracies 
\cite{nature-jarillo-2005,brokensu4,prl-groverasmussen-2012,%
prl-jarilloherrero-2005,nphys-deshpande-2008}. The gradual transition between 
transport regimes has been studied mostly via the impact of a changing gate 
voltage on a repetitive shell filling pattern in linear response
\cite{prl-makarovski-2007,prl-anders-2008,arxiv-yang-2020,prb-babic-2004,%
prb-makarovski-2007}. This has allowed to observe and model the emergence of 
the Kondo effect in its SU(2) and SU(4) manifestation and the transition 
between Coulomb blockade and Fabry-P\'erot oscillations 
\cite{prl-anders-2008,arxiv-yang-2020}. The interplay of Kondo transport, where 
on-site repulsion (i.e., charging energy) is still relevant, and quantum 
interference in an open system however still poses conceptual challenges, 
recently leading to the proposal of ``correlated Fabry-P\'erot oscillations'' in 
the transition regime \cite{arxiv-yang-2020}.

Finally, features quite similar to the diagonal lines of Fig.~\ref{fig2} can be 
found in \cite{prl-groverasmussen-2012}, see Fig.~3(c) there. The authors 
describe their observations as cotunneling corresponding to magnetic-field 
induced level crossings; a functional renormalization group calculation is used 
to successfully model many details of their measurements. Similar approaches 
may be able to cover at least part of the parameter range of our data shown in 
Fig.~\ref{fig2}.

\section{Conclusions}

We present millikelvin transport spectroscopy measurements on a 
well-characterized carbon nanotube device 
\cite{highfield,kondocharge,magdamping,heliumdamping}, where in a strong axial 
magnetic field $\bpar \le 17\un{T}$ the entire range of transport regimes from 
transparent conduction to Coulomb blockade can be traced at well-known 
absolute hole number. A multitude of quantum ground state transitions and 
Kondo-like resonances emerges in the interacting few-hole system. While similar 
phenomena have been observed and analyzed for smaller parameter spaces in 
literature, see, e.g., \cite{prl-groverasmussen-2012}, a theoretical 
description covering strong Coulomb blockade, the Kondo regime, as well as 
quantum interference in an open conductor in a consistent way is still lacking. 
Our raw data is available to the interested reader \cite{zenodo}; we hope to 
inspire corresponding work.

\begin{acknowledgements} 
The authors acknowledge funding by the Deutsche Forschungsgemeinschaft via Emmy
Noether grant Hu 1808/1, SFB 631, SFB 689, and SFB 1277, and by the 
Studienstiftung des Deutschen Volkes. A.~K.~H. acknowledges support from the 
Visiting Professor program of the Aalto University School of Science.
We would like to thank M. Grifoni, M. Marga\'nska, and P. Hakonen for 
insightful discussions, and Ch.~Strunk and D.~Weiss for the use of experimental 
facilities. The data has been recorded using Lab::Measure\-ment 
\cite{labmeasurement}.
\end{acknowledgements}

\bibliography{paper}

\end{document}